\documentclass[10pt,twoside]{IEEEtran}
\input{epsf}
\usepackage{cite}
\usepackage{amsmath} %<--pacchetto American Mathematician Society
\usepackage{amssymb} %<--pacchetto pi� ampio per simboli matematici
\usepackage{graphicx}
\input{epsf}

\renewcommand{\Pr}{\mathbb{P}}

\newcommand{\beq}{\begin{equation}}
\newcommand{\eeq}{\end{equation}}
\newcommand{\beqa}{\begin{eqnarray}}
\newcommand{\eeqa}{\end{eqnarray}}
\newcommand{\dfz}{{:=}}
\newcommand{\cH}{{{\cal H}}}
\newcommand{\test}{\mbox{$\begin{array}{c}
\stackrel{\stackrel{ \textstyle{\cal H}_1 }{\textstyle >}}
{\stackrel{\textstyle <}{\textstyle {\cal H}_0}} \end{array}$}}

\newcommand{\bx}{\mbox{\boldmath{$x$}}}

\newcommand{\cM}{{\cal M}}
\newcommand{\cPM}{{{\cal M}_n^{+}}}
\newcommand{\cNM}{{{\cal M}_n^{-}}}

\newcommand{\cPMM}{{{\cal M}_{N}^{+}}}
\newcommand{\cNMM}{{{\cal M}_{N}^{-}}}

\newcommand{\cE}{{\cal E}}

\newcommand{\wcNM}{{\widetilde{\cal M}_n^{-}}}

\newcommand{\U}{{\cal U}}

\newcommand{\E}{\mathbb{E}}

\begin{document}
%\title{Distributed Detection in Large Sensor Networks with One-Bit Communication}
%\title{One-Bit Distributed Detection in Large Sensor Networks}
%\title{One-Bit Distributed Detection in Large Wireless Networks}
%\title{Consistency of One-Bit Distributed Detection in Multisensor Systems}
\title{Asymptotic Properties of One-Bit Distributed Detection with Ordered Transmissions}

\author{Paolo~Braca, Stefano~Marano, and Vincenzo~Matta
\thanks{The authors are with the Department of Electronic and Computer Engineering (DIEII), University of Salerno,
via Ponte don Melillo I-84084, Fisciano (SA), Italy. E-mails: \{pbraca, marano, vmatta\}@unisa.it.}}

\maketitle

\begin{abstract}

Consider a sensor network made of remote nodes connected to a common fusion center. In a recent work Blum and Sadler~\cite{BlumSadlerSP08} propose the idea of ordered transmissions ---sensors with more informative samples deliver their messages first--- and prove that optimal detection performance can be achieved using only a subset of the total messages. Taking to one extreme this approach, we show that just a single delivering allows making the detection errors as small as desired, for a sufficiently large network size: a one-bit detection scheme can be asymptotically consistent. 

The transmission ordering is based on the modulus of some local statistic (MO system). We derive analytical results proving the asymptotic consistency and, for the particular case that the local statistic is the log-likelihood ($\ell$-MO system), we also obtain a bound on the error convergence rate. All the theorems are proved under the general setup of random number of sensors.
Computer experiments corroborate the analysis and address typical examples of applications including: non-homogeneous Poisson-deployed networks, detection by per-sensor censoring, monitoring of energy-constrained phenomenon. 

%One extreme way of saving energy for communication in large wireless systems designed for detection purposes is that of communicating just one bit. Such bit could be a local decision made by one sensor that, hopefully, has made the best measurement of the surrounding environment. Elaborating on a work by Blum and Sadler~\cite{BlumSadlerSP08}, we propose a wireless sensor network with these characteristics. 
%
%After sensing the surrounding environment, a data fusion stage starts in which sensors attempt to deliver their local decisions about a binary state of the nature. The time instant of such delivering is selected on the basis of the modulus of the local statistics (MO system), or on the basis of the modulus of the local log-likelihood ($\ell$-MO); the rationale is: the more the observation is informative, the sooner the decision is delivered.
%We investigate the asymptotic detection performance of a systems where only the first delivering is retained, showing the consistency of the test and deriving, for the $\ell$-MO network, bounds on the convergence rate of the error probabilities. 
%Computer experiments corroborate the theoretical findings and address some practical issues not highlighted by the asymptotic analysis. 

\end{abstract}

%%%%%%%%%%%%%%%%%%%%%%%%%%%%%%%%%%%%%%%%%%%%%%%%%%%%%%%%%%%%%%%%%%%%%%%%%%%%%
\section{Introduction}
%%%%%%%%%%%%%%%%%%%%%%%%%%%%%%%%%%%%%%%%%%%%%%%%%%%%%%%%%%%%%%%%%%%%%%%%%%%%%
\PARstart{F}{ollowing} a general trend in the area of signal processing (see e.g.,~\cite{SPMAGAZINE-special,chong&kumar}), 
in the last few decades there has been a considerable interest in distributed detection systems where a multitude of small sensors, properly networked to operate as a whole, takes the place of a single complex device typical of the classical system architecture. 

There are several advantages that the distributed schemes boast about, in comparison with their centralized counterpart. These include robustness, scalability, flexibility, portability, failure resilience, and so forth. But the advent of distributed systems also poses new challenges to the signal processing community, since the detection layer is interleaved with the communication one so that new design trade-offs arise, yielding novel design guidelines and approaches~\cite{akyildiz-survey,chamberland,viswanathan97,blum97}. 

One aspect that is of primary relevance in the implementation of many distributed detection systems is the limitation of the sensors' energy, with consequence on the sensors' capability of sensing, processing and delivering data. While a precise evaluation of the relative impact of these tasks strongly depends upon the specific network, for many wireless systems the task of communication is by far the more energy consuming~\cite{sadler-tutorial,appadwedulaJSAC}. As a consequence, one important issue in a wireless sensor network (WSN) is how to design the system in order to reduce the amount of communication, given a desired level of detection performance. 

As regards to this aspect, in the literature several approaches have been proposed. They include the design of energy efficient routing in ad-hoc networks~\cite{akyildiz-survey}, the implementation of proper strategies for the access to the common communication medium~\cite{akyildiz-survey}, and the use of censoring pioneered in~\cite{rago-censoring} and further developed in many successive works, see e.g.,~\cite{tsitsikliscensoring,appadwedulaSP,addesso-censoring-SP07}. Censoring refers to the idea of quantifying, at the sensor level, the informativeness of the sensed samples before sharing the observations with other nodes or with a sink unit: only the data that are believed informative enough are shared, while the node remains silent otherwise, optimizing the battery life.

In this work we consider a modification of this idea: we design a distributed detection system for binary hypothesis test in which the remote nodes quantify the informativeness (for detection) of their observations, and communicate their local decisions to the system, with more informative samples that are communicated first.
As soon as one local decision is communicated, that is taken as the global decision of the network and the detection task terminates. Thus, censoring is obtained on a time-selective basis.

Note that two different stages exists: the sensing stage in which all the nodes observe the state of the nature to be decided over, and the successive data fusion stage in which each node is ready to deliver its local decision after a time interval that is inversely proportional to the informativeness of the observations. As the first (more informative) local decision is sent, the whole detection task is terminated with final decision equal to the local quickest decision. To make a decision, our scheme prescribes just one communication event.

A WSN can be organized according to many different architectures. One possibility is that a common fusion center (FC) exists with the role of collecting the data that the remote nodes deliver, usually after some local pre-processing. In this case, either dedicate links connecting each node to the FC exist, or there is a common channel that the sensors access by some suitable multiple access scheme.

Another common architecture lacks any central unit, and the local decisions are propagated within the network by inter-sensor communication protocols, while collaborative signal processing procedures are employed to mimic the presence of a FC. In these ``fully flat'' WSNs, the final decision is taken by the network in a distributed fashion and is usually shared by all the nodes, as in the consensus schemes~\cite{braca-spawc09,asymptotic-rc,fusion2008,running-cons,scaglione-varshney}.

For most part of this paper, we do not refer to any specific architecture since our goal is to investigate the detection performance of the statistics computed at the faster \emph{firing} node of the network, which is largely independent of the specific WSN architecture. For concreteness, one can imagine that, if a FC exists, after receiving the first delivering from some sensor, such FC broadcasts a stopping message to all the other nodes. Conversely, in fully flat architectures, the halting command should be propagated, along with the decision, from the deciding node to all the other nodes by means of some suitable multi-hop protocol or by means of consensus algorithms. 
We refer to our scheme as a one-bit distributed detection; it should be noted, however, that we are disregarding the network messages required for the halting procedure.

\subsection{Related work \& motivations}

The general approach of quickly computing an efficient detection statistic resembles the quickest detection problem originally introduced by~\cite{Page}; see~\cite{basseville-book,poorbook-quickest,Bracaetal-Pageconsensus} for more recent references.
The substantial difference is that quickest detection procedures are usually used to monitor in a continuous way an underlying phenomenon to discover a change in the statistics of the observed process. Conversely, in the system we design, the sensing stage and the detection stage are separated. First, the environment is sensed by all the sensors, and then a detection step is initiated. 
The quickness is only a property of this second stage, it is not related to event detections but it is a mean to save system energy, and there is no change in the process statistics. %As a consequence, our performance indices, and the general approach, are different from those of quickest detection literature. 

More relevant to our setup is the work by Blum and Sadler~\cite{BlumSadlerSP08}. They study a WSN engaged in a detection problem and conceive a multiple access architecture in which each sensor accesses the channel after a delay inversely proportional to the informativeness of its measurement. 
They show that, upon receiving at the FC a certain fraction of the overall sensor measurements, a decision can be made at the same performance level achievable by using the complete set of data collected by the sensors. The channel access rule, indeed, is such that the more informative samples are delivered first so that transmissions can be saved, without degradation in error probability. 

In applications where sensors are severely battery-limited and relatively tiny and cheap, one can lead this approach to one extreme. What if, rather than collecting at the FC a certain fraction of the sensors' deliverings, just one single sample is considered? This would imply a significant energy saving, payed in the coin of a performance degradation with respect to the approach of~\cite{BlumSadlerSP08}. However, given that sensors are tiny and cheap, performances can be improved by increasing the number of sensors, if some form of asymptotic consistency holds. This work elaborates on this concept.

%A standard asymptotic settings for detection~\cite{Lehmann-testing} is based on the asymptotically normal models, i.e., on the large-sample properties of the {\em sums} of random variables~\cite{Khintchine}. As argued in~\cite{Gnedenko}, many analogies exist between these sums and the asymptotic properties of the {\em order statistics} of a set of random variables, that motivated wide interest in the study of the Extreme Value Theory (EVT), see, {\em e.g.},~\cite{Gnedenko, Galambos}. This motivates our asymptotic setting for hypothesis testing via order statistics.

\subsection{Main results \& organization}
\label{subsec:mainres}

One main theoretical result of this paper is the proof of the asymptotic consistency of the described one-bit system when the informativeness of sensors' samples is evaluated according to the modulus $|T(\cdot)|$ of some suitable transformations $T(\cdot)$ of the observed samples. For instance, in the case that $T(\cdot)$ is the identity, the idea is that extreme values of the measurements carry more information for detection with respect to ``near the mean'' observations. 

We also consider as index of informativeness the modulus of the log-likelihood of the observed sample, which is motivated by known results on censoring~\cite{rago-censoring}. When remote sensors compute the log-likelihoods, and the delivering time is accordingly set, beside proving the asymptotic consistency of the test we derive bounds to the asymptotic rate of convergence of the error probabilities. 
We consider also networks whose size is random and possibly depends upon the observed data. This allows to consider very general applicative scenarios, examples of which are given in Sect.~\ref{sec:appl}.

The remainder of this paper is organized as follows. The problem statement is described in Sect.~\ref{sec:state}, the main results are presented in Sect.~\ref{sec:theory}, examples of applications are provided in Sect.~\ref{sec:appl}, while
in Sect.~\ref{sec:concl} we summarize.

\section{Problem statement}
\label{sec:state}

\subsection{Preliminaries}

Consider a WSN made of $n$ remote units that sense the surrounding environment to decide which of two mutually exclusive states of the nature, ${\cal H}_0$ or ${\cal H}_1$, is actually in force. The observation made by sensor $i$ is modeled as a random variable $X_i$, where $i=1,2,\dots, n$, and the $X_i$'s are independent and identically distributed (iid) samples drawn from one of the two possible marginal probability density functions (or pdf's) $f_X(x;\cH_j)$, $j=0,1$. This simple hypothesis test can be schematically formalized as
\beq
{\cal H}_0: \; X_i \sim f_X(x;\cH_0), \quad \textnormal{vs.} \quad 
{\cal H}_1: \; X_i \sim f_X(x;\cH_1). \label{eq:test}
\eeq
We assume throughout this work that the involved random variables, taking values in $\Re$, admit densities and these densities have 
unbounded support, in the sense that $\sup_x \{f_X(x)>0\} = \infty$ and
$\inf_x \{f_X(x)>0\} = -\infty$. 

Suppose that sensor $i$ computes a suitable local detection statistic $T(X_i)$, to be compared with a certain threshold value\footnote{Needless to say, a likelihood ratio test would be the best. However, this might not be available, e.g., in fully or partially nonparametric setups, such that it is of interest to study general detection statistics, see also~\cite{kassam}.}. If the threshold is crossed a local decision in favor of ${\cal H}_1$ is made, while the local decision is for ${\cal H}_0$ otherwise; let $D_i=0,1$ be such decision.
As proposed in~\cite{BlumSadlerSP08}, sensor $i$ is programmed to communicate with the network after a time interval proportional to $1/|T(X_i)|$, however in this work it is supposed that sensor $i$ delivers his local decision $D_i$ instead of the local statistic $T(X_i)$ as in~\cite{BlumSadlerSP08}. We assume that the sensors are perfectly  synchronized so that they share the same time reference. Then, the larger is $|T(X_i)|$, the faster is the delivering of $D_i$ and, different from~\cite{BlumSadlerSP08}, in our scheme the ``winner takes all''. 
Otherwise stated, as soon as the quickest sensor delivers its own decision (say, the sensor ``fires''), such decision is immediately taken as the final one for the whole system, all other transmissions by the remaining $n-1$ sensors are instantaneously inhibited, and the overall detection process is terminated.

While many other forms of ordering are certainly conceivable, the choice of modulus ordering leads to analytical tractability and has a precise rationale, as detailed later. Two obvious choices for the transformation $T(\cdot)$ are the identity $T(x)=x$, and that based on the log-likelihood ratio
\[
T(x)=L(x)\dfz\log \frac{f_X(x;\cH_1)}{f_X(x;\cH_0)}.
\]
In the former case the firing time of the generic sensor $i$ is proportional to $1/|X_i|$, in the latter it is proportional to  $1/\left|L(X_i)\right|$.

The idea of accessing the channel by ordering is borrowed by~\cite{BlumSadlerSP08}, and the main aim of this paper is to investigate the asymptotic properties of the above distributed detector, with respect to the number $n$ of sensors. 
A number of simplifying assumptions are made, including the possibility of instantaneously communicate the first local decision to sleep down the system, and the assumption of perfect time synchronism among sensors. 
While we use this setup to get clean analytical results and useful insights, some of the effects related to time errors and uncertainty are briefly investigated in Sect.~\ref{sec:offset}, exhibiting a certain robustness of the proposed strategy.
%Without these abstractions, the addressed problem would be much more difficult to be investigated and the results would lead to less intuitive understanding. 

\subsection{Detector design}

First, let us specify the local testing rule of~(\ref{eq:test}) at sensor~$i$ 
\beq
%T\left(X_i\right) \overset{\cH_1}{\underset{\cH_0}{\gtrless }} \gamma_n,
T\left(X_i\right) \test \gamma_n,
\label{eq:test0}
\eeq  
where $\gamma_n$ is the detection threshold (which is allowed to depend on $n$), and the local decision $D_i$ is accordingly defined.  

The transmission policies considered in this work is defined as follows.

\vspace*{5pt} \noindent
\textsc{Definition 1} {\em (Transmission policy) 
The transmission of the local decision $D_i$ made by the generic sensor~$i$ is activated at a time inversely proportional to the absolute value of its transformed measurement $|T\left(X_i\right)|$; we call this policy MO (modulus ordered). 
Within the class of MO, if $T(x)=L(x)$ the system is called $\ell$-MO (log-likelihood modulus ordered).}~\hfill$\diamond$

\vspace*{2pt} \noindent 
Thus, the transmission policy is identified by the transformation $T(\cdot)$, leading to the definition of the random variable $Z_i := T(X_i)$, with cumulative distribution function (cdf) $F_{Z}\left(x;\cH_j\right)$ and pdf $f_{Z}\left(x;\cH_j\right)$, under hypothesis $\cH_j$ with $j=0,1$. The modulus ordering can be defined in terms of the index permutation $\pi(\cdot)$ defined by the property that
\beq
\left| Z_{\pi(1)} \right|\leq \left| Z_{\pi(2)} \right| \leq\dots\leq \left| Z_{\pi(n)} \right|,
\eeq
and the decision statistic of our system is $\cM_n := Z_{\pi(n)}$. Therefore, the decision rule of the test~(\ref{eq:test}) for the whole network is:
\beq
%\cM_n \overset{\cH_1}{\underset{\cH_0}{\gtrless }} \gamma_n.
\cM_n \test \gamma_n.
\label{eq:test1}
\eeq  
%where $\gamma_n$ is the detection threshold, which is allowed to depend on $n$. 

Next, consider the following extreme value statistics ($k$ is a positive integer) %\footnote{Note that $\cNM=-\min(Z_1, \dots, Z_k,\dots,Z_n)$.}
%\beqa
%\cPM &\dfz& \max(Z_1, \dots, Z_k,\dots,Z_n), \nonumber \\
%\cNM &\dfz& \max(-Z_1,\dots,-Z_k,\dots,-Z_n), \nonumber
%\eeqa
\[
\cPM \dfz \max_{k\leq n} Z_k,\qquad \cNM \dfz - \min_{k\leq n} Z_k=\max_{k\leq n} 	\left(-Z_k\right),
\]
whence the decision statistic in~(\ref{eq:test1}) can be expressed as 
\beq
 \cM_n = \left\{\begin{array}{rcr} \cPM & \textnormal{if} & \cPM \geq \cNM \\ -\cNM & \textnormal{if} & \cPM < \cNM \end{array}\right. .
 \label{eq:M_N}
\eeq
Also, let us denote by $F_{\cM_n}\left(x;\cH_j\right)$, $F_{\cPM}\left(x;\cH_j\right)$ and $F_{\cNM}\left(x;\cH_j\right)$ the cdf's of the above quantities under hypothesis $\cH_j$, and by $f_{\cM_n}\left(x;\cH_j\right)$, $f_{\cPM}\left(x;\cH_j\right)$ and $f_{\cNM}\left(x;\cH_j\right)$ the corresponding pdf's. Standard results of order statistics theory allows us to compute these functions as follows~\cite{Arnold}:
\begin{align}
&f_{\cPM}\left(x;\cH_j\right) = n\,F_Z^{n-1}\left(x;\cH_j\right)f_Z\left(x;\cH_j\right) 
\label{eq:PM_pdf}  \\  
&f_{\cNM}\left(x;\cH_j\right) = n \left(1-F_Z\left(-x;\cH_j\right)\right)^{n-1}f_Z\left(-x;\cH_j\right) . %\nonumber
\label{eq:NM_pdf}
\end{align}

As regard to the statistical distribution of $\cM_n$, exploiting the results provided in~\cite{Bairamov97,Arnold_multivariate} we have
\beq
f_{\cM_n}(x;\cH_j) = n\, h_j^{n-1}(x)\,f_{Z}(x;\cH_j) 
\label{eq:MN_pdf}
\eeq
where 
\beq
h_j(x) = F_{Z}(|x|;\cH_j)-F_{Z}(-|x|;\cH_j).
\label{eq:h}
\eeq

%that, exploiting~(\ref{eq:MN_pdf}) and~(\ref{eq:h}), can be also written
%\beqa
%\alpha_n = \displaystyle {n\,\int^{\infty}_{\gamma_n}  h_0^{n-1}(x) \,f_{Z}(x;\cH_0) dx},
%\label{eq:alpha} \\
%\beta_n = \displaystyle{n\,\int^{\gamma_n}_{-\infty}  h_1^{n-1}(x) \,f_{Z}(x;\cH_1) dx}.
%\label{eq:beta}
%\eeqa

\section{Asymptotic analysis}
\label{sec:theory}

We are now ready to introduce the considered asymptotic setup for order statistics. 
We are primarily interested in the regime of large number of sensors, that is, $n\rightarrow\infty$. 
However, in many WSN applications, the number of effective sensors that contribute to the final inference is uncertain, due to several practical issues, such as failures, time-varying topologies, impaired communication, compromised nodes, and so on. 
Accordingly in this paper we consider the more general case of a random number of sensors; see, e.g.,~\cite{Niu&Varshney}, for a discussion on the relevance of this scenario in distributed detection problems.

Formally, let $N$ be the random number of sensors, whose distribution depends on an integer parameter\footnote{Here we formally consider $\nu$ as an integer parameter, thus the $\nu^{th}$ element of a generic sequence $\{a_n\}_{n=1}^{\infty}$ can be denoted as $a_{\nu}$. However, all the asymptotic results of this work also hold when $\nu$ is real, and $a_{\nu}$ is replaced by $a_{\lfloor\nu\rfloor}$.} $\nu$. 
Depending on the application, $\nu$ may represent the total number of available sensors ($N$ of which are in fact activated), the (integer part of the) expected number of sensors $\nu=\lfloor\E(N)\rfloor$, and so forth.
In order to define a proper asymptotic setup, the precise sense in which the random $N$ diverges must be defined. A general and convenient  formalization is to assume that, as the parameter $\nu$ goes to infinity
\beq
\frac{N}{\nu}\rightarrow R\quad\textnormal{in probability},
\label{eq:Nrand}
\eeq
where $R$ is a positive random variable, see~\cite{Galambos}.

\subsection{Relevant EVT background}
Before illustrating the main asymptotic theorems, we briefly summarize some relevant facts from the classical literature.
Let $Y_1,Y_2,\dots,Y_n$ be a collection of iid random variables with unbounded support, and let $M_n=\max_{k\leq n} Y_k$.
%A classical result of the EVT is that, provided that suitable regularity conditions are met, the maximum $M_n=\max_{i\leq n} Y_i$ belongs to one among three universal domains of attraction:   

\vspace*{5pt} \noindent
\textsc{Lemma 1} {\em (Attraction) 
Under mild regularity conditions, there exist sequences of normalizing constants $a_n$, $b_n$ such that
\beqa
\lim_{n\rightarrow \infty} F_{M_n}(a_n\,x+b_n) = G(x) 
\eeqa
where $G(x)$ is either the Gumbel distribution or the Fr\'echet distribution.
}
~\hfill$\diamond$ 

\vspace*{2pt} \noindent
The technical regular conditions can be found in any textbook on EVT (e.g.,~\cite{Gnedenko}) and the relevant features of the quoted distributions are as follows~\cite{Gnedenko}:\footnote{To avoid confusion, notice that the assumption of unbounded support rules out convergence to the third class of attraction, namely the Weibull distribution.}

\vspace*{5pt}
\noindent
{\textsc{Gumbel}}
\vspace*{-5pt}
\beqa
&&G(x)=e^{-e^{-x}},\quad -\infty<x<\infty,\nonumber\\ 
&&b_n=F_Y^{-1}\left(1-\frac{1}{n} \right), ~~a_n = \frac{1}{n\,f_Y (b_n )} ,\nonumber\\
&&\lim_{n\rightarrow\infty}(b_n/a_n)=\infty.
\nonumber
\eeqa
%\vspace*{5pt}
\noindent
{\textsc{Fr\'echet}}
\vspace*{-5pt}
\beqa
&&G(x)=
\left\{\begin{array}{ll} 0 & x\leq 0 \\\exp(-x^{-\xi}) & x>0, \end{array}\right.\quad \xi>0,
\nonumber\\
&&b_n=0,~~a_n = F_Y^{-1}\left(1-\frac{1}{n} \right),\nonumber\\ 
&&\lim_{n\rightarrow\infty}a_n=\infty.
\nonumber
\eeqa

An extension of Lemma~1 to the case of random number of variables has been proved by Galambos~\cite{Galambos}: 

\vspace*{5pt}
\noindent{\textsc{Lemma 2}} {\em (Attraction with random number of variables) Let $N$ be an integer random variable and let, as $\nu\rightarrow\infty$, $N/\nu$ converge in probability to a positive random variable $R$. If
\beq
\lim_{n\rightarrow \infty} F_{M_n}(a_n\,x+b_n) = G(x),
\eeq
then
\beq
\lim_{\nu\rightarrow \infty} F_{M_N}(a_{\nu}\,x+b_{\nu}) = \E\left(G^R(x)\right)
\label{eq:weakrandom}
\eeq
where the expectation is taken under the distribution of $R$.~\hfill$\diamond$ 
}

\vspace*{5pt} \noindent \textsc{Definition 2} {\em (Right/Left tail dominance) 
Given a random variable $Y$, consider the limit 
\beq
\lim_{x\rightarrow\infty} \frac{1-F_Y(x)}{F_Y(-x)}
\label{eq:righttail}
\eeq
We say  that $Y$ is right-tail dominant if the above limit is $+ \infty$, and we say that $Y$ is left-tail dominant if the limit is zero.}~\hfill$\diamond$

\subsection{Detection asymptotic properties} 
 
\begin{figure}
\centerline{\leavevmode \epsfxsize=3.5in %\epsfysize=1.2in 
\epsfbox{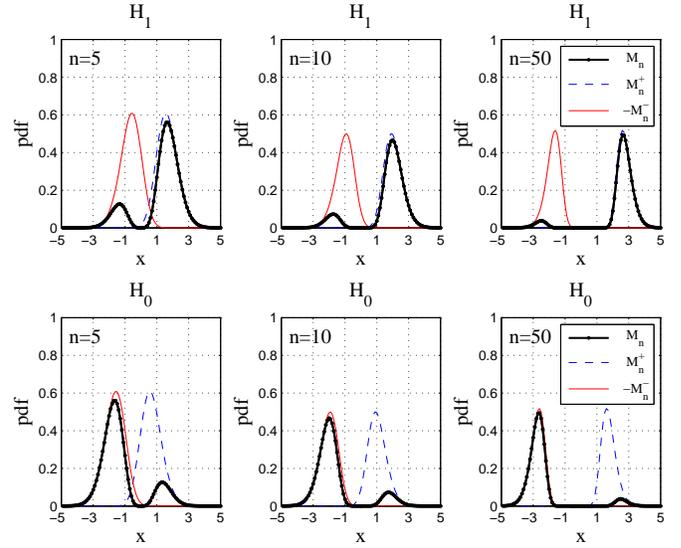}
}
\caption{Illustrative sketch of the intuition behind this work: Under ${\cal H}_1$, the pdf of $\cM_n$ is closer and closer to that of $\cPM$, as $n$ grows, and the small peak located around the peak of the pdf of $-\cNM$ tends to vanish. 
Conversely, under ${\cal H}_0$, the pdf of $\cM_n$ approaches that of $-\cNM$. }
\label{fig:rationale}
\end{figure}

Let us explain the rationale of the modulus ordering, and for sake of simplicity assume the case in which the number of sensor is deterministic $N=n$. Suppose $n$ ``very large'', local decision of the firing sensor~(\ref{eq:test1}) is based either on the largest $\cPM$ or on the smallest $-\cNM$ of the $n$ transformed samples collected by the system. However, if the right tail of the distribution $f_Z(x;\cH_1)$ dominates over the left tail (i.e., the right tail is heavier, or decreases slower) then with high probability the local decision is made using the largest transformed sample collected by the network. The converse happens with left-tail dominant distribution. See Fig.~\ref{fig:rationale} for an instance of this effect.

In practical problems it is often the case that $f_Z(z;\cH_0)$ is left dominant while $f_Z(z;\cH_1)$ is right dominant (or viceversa), so that the hypothesis test can be thought as one comparing a very large positive sample against a very small negative value. Based on this argument, one expects that the error probability can be made smaller and smaller as $n$ grows. Here below this intuition is verified and the sense in which the errors can be controlled is made precise.

\vspace*{5pt} \noindent
\textsc{Theorem 1} 
{\em (Asymptotics of MO detection statistic) 
Consider an MO network with a random number of active sensors $N$. Suppose that 
\begin{itemize}
\item[{\em i)}]
The random variables $Z_1,Z_2,\dots$ have unbounded support.
\item[{\em ii)}]
$\cPM$ is attracted under $\cH_1$ with normalizing constants $a_n^+$ and $b_n^+$, and limiting distribution $G^+(x)$. 
Similarly $\cNM$ is attracted under $\cH_0$ with normalizing constants $a_n^-$ and $b_n^-$, and limiting distribution $G^-(x)$. 
\item[{\em iii)}]
The ratio $N/\nu$ converges to a positive random variable $R_1$ under $\cH_1$, and to a positive random variable $R_0$ under $\cH_0$.
\item[{\em iv)}]
The random variables $Z_i$'s are right-tail dominant under $\cH_1$ and left-tail dominant under $\cH_0$. 
\end{itemize}
Then 
%\beq
%\frac{\cM_N}{\cPMM}\rightarrow 1 \textnormal{ under $\cH_1$ }
%\qquad
%\frac{\cM_N}{\cNMM}\rightarrow -1 \textnormal{ under $\cH_0$ }
%\label{eq:mainthRATIO}
%\eeq
%and 
%\[
%\cM_N -\cPMM \rightarrow 0\textnormal{ under $\cH_1$ },
%\;
%\cM_N-\cNMM \rightarrow 0 \textnormal{ under $\cH_0$ },
%%\label{eq:mainthDIFF}
%\]
\beq
\frac{\cM_N}{\cPMM}\rightarrow 1, 
\quad
\cM_N -\cPMM \rightarrow 0,\quad\textnormal{ under $\cH_1$,}
\label{eq:mainth_H1}
\eeq
and 
\beq
\frac{\cM_N}{\cNMM}\rightarrow -1,
\quad
\cM_N+\cNMM \rightarrow 0,\quad \textnormal{ under $\cH_0$,}
\label{eq:mainth_H0}
\eeq
all the convergences being in probability.}~\hfill$\diamond$ 

\vspace*{2pt} \noindent
{\em Proof:} The proof is deferred to Appendix~\ref{Sec:TH1}. 

\vspace*{2pt}
We want to stress that conditions $i)$, $ii)$ and $iv)$ are by no means restrictive and hold true in a large number of practical applications. Condition $iii)$ is a convenient way to handle with networks of random size and encompasses as special case the scenario of nonrandom $N$. The claim of the theorem, in words, states that the detection statistic $\cM_{N}$ tends (asymptotically) to behave like $\cPMM$ under $\cH_1$, and like $-\cNMM$ under $\cH_0$, see again Fig.~\ref{fig:rationale}.
As a direct consequence of Theorem~1, we get the following.
%\newpage

\vspace*{5pt} \noindent
\textsc{Corollary} {\em (Attraction of the detection statistic) Under the same assumptions of Theorem 1
%\vspace*{-5pt}
\beq
\lim_{\nu\rightarrow \infty} F_{\cM_{N}}\left(a^{+}_{\nu}\,x+b^{+}_{\nu}; \cH_1\right) = H^+(x), 
\label{eq:mainth1}
\eeq
and 
\vspace*{-5pt}
\beq
\lim_{\nu\rightarrow \infty} F_{\cM_{N}}\left(a^{-}_{\nu}\,x - b^{-}_{\nu}; \cH_0\right) = H^-(x)  
\label{eq:mainth2}
\eeq
where 
\vspace*{-10pt}
\beqa
H^+(x)&=&\E\left(\left(G^+(x)\right)^{R_1}\right), \\ 
H^-(x)&=&1-\E\left(\left(G^-(-x)\right)^{R_0}\right). 
\label{eq:Hpmdef}
\eeqa}~\hfill$\diamond$

\vspace*{2pt} \noindent
{\em Proof:} Assume that $\cH_1$ is in force. In view of Theorem~1 it follows that $\left(\cM_N-b^+_{\nu}\right)/a^+_{\nu}$ converges in probability to $\left(\cPMM-b^+_{\nu}\right)/a^+_{\nu}$. Assumption $ii)$, along with a direct application of Lemma~2 implies the convergence in distribution of $\left(\cPMM-b^+_{\nu}\right)/a^+_{\nu}$. Then, the convergence in distribution of $\left(\cM_N-b^+_{\nu}\right)/a^+_{\nu}$ claimed in~(\ref{eq:mainth1}) follows by a direct application of Theorem~2.7 in~\cite{vandervaart}.
The proof for $\cH_0$ is similar.~\hfill$\bullet$  

As to the performance of the hypothesis test, this is expressed in terms of the false alarm and miss detection probabilities
\beqa
\alpha_{\nu}=\Pr ( \textnormal{decide}\,\cH_1 ;\cH_0 ) = \Pr (\cM_N \geq \gamma_{\nu};\cH_0 ), 
\label{eq:alpha0} \\
\beta_{\nu}=\Pr ( \textnormal{decide}\,\cH_0  ;\cH_1 ) = \Pr (\cM_N < \gamma_{\nu}  ;\cH_1 ). 
\label{eq:beta0}
\eeqa 
We note explicitly that, being $N$ random, the threshold of the test cannot be set as a function of that, but rather it must be controlled by the parameter $\nu$, that is clearly assumed known in order to fix the threshold value. 

We consider the classical setup where a prescribed (asymptotic) false-alarm level $\alpha$ is imposed, while it is required that $\beta_{\nu}$ vanishes with increasing $\nu$. 
In the light of Theorem~1, it is reasonable to impose an asymptotic false-alarm level $\alpha$ based on the asymptotic distribution under $\cH_0$.
We indeed know that $(\cM_N+b^-_\nu)/a^-_\nu$ converges in distribution toward $H^-(x)$. This implies that the threshold $\gamma_\nu=a^-_\nu\gamma-b^-_\nu$, with
\[
\alpha=1-H^{-}(\gamma)=\E\left(\left(G^-(-\gamma)\right)^{R_0}\right),
\]
achieves the asymptotic false alarm $\alpha$.

For this computation, it is useful to define the false alarm $\widetilde\alpha$ corresponding to a system with deterministic number of sensors, that is
\beq
G^-(-\gamma)=\widetilde\alpha
~\Longrightarrow~
\gamma=\left\{
\begin{array}{lr}
\log\log(1/\widetilde\alpha)&\textnormal{GUMBEL}
\\
-\left(\log(1/\widetilde\alpha)\right)^{-\frac{1}{\xi}}&\textnormal{FR\'ECHET}
\end{array}
\right.
\label{eq:alphadet}
\eeq
The required false alarm $\alpha$ can be computed as a function of $\widetilde\alpha$, by \beq
\widetilde\alpha:\quad \E\left(\widetilde\alpha^{R_0}\right)=\alpha.
\label{eq:inversion}
\eeq

The threshold $\gamma_{\nu}=a^-_\nu\gamma-b^-_\nu$ is selected by using the asymptotic $\cH_0$-distribution $H^-(x)$.
An alternative might be that of imposing the strict equality $\alpha_{\nu}=\alpha$ for any finite $\nu$.
This, however, would require exact knowledge of the cdf of the detection statistic for any finite $\nu$, which is usually unavailable.

On the other hand, it is possible to use the asymptotic ``similarity'' (under $\cH_0$) between $\cM_N$ and $-\cNMM$ to set a new threshold  
as $(1-F_Z\left(\gamma_\nu;\cH_0\right))^\nu=\widetilde\alpha$, which can be shown to achieve asymptotically the desired false-alarm level.
These results are summarized in the following theorem.

\vspace*{5pt} \noindent
\textsc{Theorem 2} 
{\em (MO consistency)
Under the assumptions of Theorem~1: \newline
$i)$ The nonparametric setting $\gamma_\nu=0$ ensures that 
\beq
\alpha_\nu+\beta_\nu\rightarrow 0.
\label{eq:nonparam}
\eeq
$ii)$ Let the detection threshold be either
\beq
\gamma_{\nu}=a^-_{\nu} \gamma - b^-_{\nu}, 
\label{eq:gamma}
\eeq
where $\gamma$ solves $1-H^-(\gamma)=\alpha$, or
\beq
\gamma_{\nu}=F_Z^{-1}\left(1-\widetilde\alpha^{\frac{1}{\nu}};\cH_0\right),
\label{eq:gammaprime}
\eeq
where $\widetilde\alpha$ solves $\E\left(\widetilde\alpha^{R_0}\right)=\alpha$.
%or
%\beq
%\gamma_{\nu}=F_Z^{-1}\left(1-\widetilde\alpha^{\frac{1}{\nu}};\cH_0\right).
%\label{eq:gammaprime}
%\eeq 
%where $\widetilde\alpha$ solves $\E\left(\widetilde\alpha^{R_0}\right)=\alpha$.
Then:
%\vspace*{-5pt}
\beq
\quad \alpha_{\nu}\rightarrow \alpha, \quad \textnormal{\emph{and}} \quad
\beta_{\nu}\rightarrow 0.
\label{eq:param}
\eeq
}~\hfill$\diamond$ 

\vspace*{2pt} \noindent
{\em Proof:} The proof is deferred to Appendix~\ref{Sec:TH2}. 

\vspace*{5pt} \noindent
{\em Remark.} The threshold setting used in eq.~(\ref{eq:nonparam}) does not require any a-priori knowledge of the statistics, that amounts to a {\em nonparametric} threshold setting. 
This may be convenient in practical applications where limited knowledge of the statistical model is available to the remote nodes.

\vspace*{2pt} 
The above theorems are valid for a general MO network. 
For specific detection problems and/or local transformations, more powerful results might be obtained.  
This is the case of an $\ell$-MO strategy (namely, when $Z_i$ is the log-likelihood of the observations) applied to the shift-in-mean problems of the kind
\beq
f_X(x;\cH_0)=\phi(x+\theta_0),\quad f_X(x ;\cH_1)=\phi(x-\theta_1)
\label{eq:shift}
\eeq 
where $\phi(x)$ is an even function, $\phi(x) > 0$ $\forall x$, $\theta_0\geq0$ and $\theta_1>0$. 
For this scenario we prove the following

\vspace*{5pt} \noindent
\textsc{Theorem 3} 
{\em ($\ell$-MO properties) 
Assume that the local log-likelihoods fulfill conditions $i)$, $ii)$ and $iii)$ of Theorem~1. 
Then, condition $iv)$ is automatically verified, and the results of Theorems~1 and~2 apply.
In addition, if $N$ is independent of the observations,  the following upper bound on the miss detection probability holds
\beq
\beta_{\nu}\leq e^{\gamma_{\nu}}.
\label{eq:Chernoff}
\eeq
}
~\hfill$\diamond$ 

\vspace*{2pt} \noindent
{\em Proof:} The proof is deferred to Appendix~\ref{Sec:TH3}.

\section{Applications}
\label{sec:appl}

To illustrate the above results, we now focus on sensor network applications. 
Both MO and $\ell$-MO systems are investigated for different case studies, with the twofold goal of 
providing a numerical check for the asymptotic convergence claimed in the theoretical results, and of investigating the effect of a moderately small number of sensors. We also consider networks of random size and a typical application example from the distributed detection domain, such as censoring sensor systems. Finally we address the case where the random network size depends upon the observations, and we briefly touch upon the robustness of the detection system to timing offsets. 

\subsection{Gaussian observations}
\label{sec:gauss}

\begin{figure}
\centerline{\leavevmode \epsfxsize=3.5in %\epsfysize=1.2in 
%\epsfbox{Gauss_gamma=0.eps}
\epsfbox{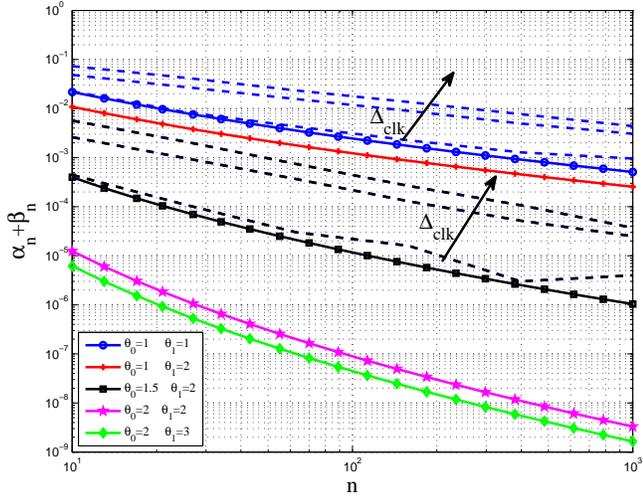}
}
\caption{MO transmission policy. Error $\alpha_n+\beta_n$ for the Gaussian nonparametric example, with $\sigma=1$ and different combinations of $\theta_0$ and $\theta_1$. The solid curves are obtained by numerical integration based on eq.~(\ref{eq:MN_pdf}). Dashed lines refer to clock offset discussed later, in Sect.~\ref{sec:offset}: It is shown the effect of synchronism errors for $\Delta_{clk}=0.1,1,2$ with respect to both the nominal case $\theta_0=\theta_1=1$ and $\theta_0=1.5,\theta_1=2$.}
\label{fig:nonparametric}
\end{figure}

Let us start by considering the following Gaussian observation model (${\cal N}(a,b)$ is our shortcut for a Gaussian distribution with mean $a$ and standard deviation $b$):
\beq
{\cal H}_0: \; X_i \sim {\cal N}(-\theta_0,\sigma) , \quad \textnormal{vs.} \quad 
{\cal H}_1: \; X_i \sim {\cal N}(\theta_1,\sigma), \label{eq:gauss}
\eeq
where $\theta_0$, $\theta_1$ and $\sigma$ are positive parameters, and the number of sensors $n$ is deterministic (namely, here we set $N=\nu=n$). Let us consider first the MO policy with $T(x)=x$. According to Theorem~2, $\alpha_n \rightarrow 0$ and $\beta_n \rightarrow 0$. This is true even if $\theta_0$, $\theta_1$, $\sigma$ and $n$, are all unknown. In this case, we are faced with a fully nonparametric test in which the sensors have no knowledge of the parameters appearing in~(\ref{eq:gauss}) and they accordingly use a zero threshold, see Theorem~2, part~$i)$. 
The results are shown in Fig.~\ref{fig:nonparametric}, where the corresponding error probabilities (solid curves) have been obtained by numerical integration based on expression~(\ref{eq:MN_pdf}): as predicted both the error probabilities go to zero, with a rate that depends upon the system parameters. The dashed curves refer to the effect of clock offsets, and we comment on this later.

\begin{figure}
\centerline{\leavevmode \epsfxsize=3.5in %\epsfysize=1.2in 
\epsfbox{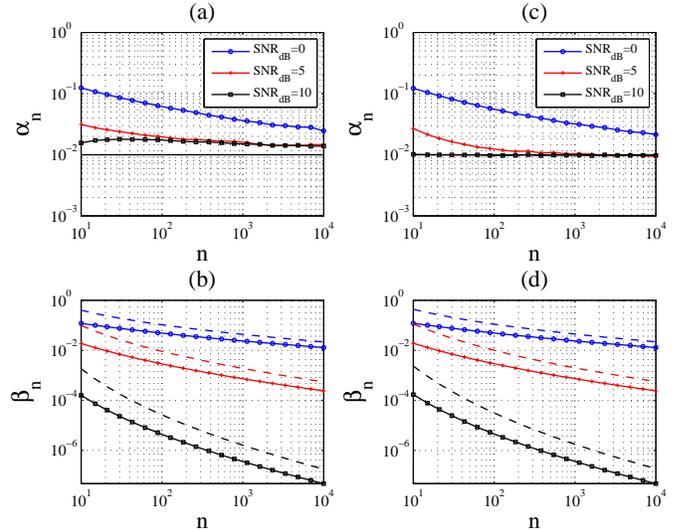}
}
\caption{$\ell$-MO transmission policy. Error probabilities $\alpha_n$ and $\beta_n$ for the Gaussian example with known parameters. The curves are obtained by numerical integration based on eq.~(\ref{eq:MN_pdf}), and are parametrized in SNR$=(\theta_1+\theta_0)/\sigma$. Panels $(a)$ and $(b)$ refer to the threshold setting given in~(\ref{eq:gamma}), while $(c)$ and $(d)$ refer to that in~(\ref{eq:gammaprime}). In the lower plots are also shown, as dashed lines, the miss detection bounds~(\ref{eq:Chernoff}).} 
\label{fig:GaussLikelihood1}
\end{figure}

With reference to the same observation model~(\ref{eq:gauss}), suppose now that the parameters are known and that the $\ell$-MO policy is in order. It is clear that the $\ell$-MO policy cannot be implemented without the knowledge of the distribution parameters, since it requires the computation of the likelihood. We assume again that $n$ is deterministic. This case lies in the application domain of Theorem~3 and the error probabilities, still computed by numerical integration based on~(\ref{eq:MN_pdf}), are illustrated in Fig.~\ref{fig:GaussLikelihood1}. 
It is worth noting that the asymptotic value is approached faster for larger values of SNR=$(\theta_1+\theta_0)/\sigma$. This should be expected because, as it can be easily seen, the tail dominance is ``stronger'' when the SNR grows. Panels $(a)$ and $(b)$ refer to the threshold setting given in~(\ref{eq:gamma}), while $(c)$ and $(d)$ refer to the threshold in~(\ref{eq:gammaprime}). We see that $\alpha_n$ converges to the desired asymptotic value (set to $\alpha=10^{-2}$ in the figure); however, in $(c)$ the convergence is somehow faster than that in $(a)$, suggesting that the threshold setting~(\ref{eq:gammaprime}) provides, in this example, some advantage. As claimed in Theorem~3, we see that $\beta_n \rightarrow 0$ as shown in panels $(b)$ and $(d)$. 
Note that the curves in $(b)$ and $(d)$ are very similar, which reveals that the selection of the threshold between the two alternatives, is not critical with respect to $\beta_n$.
Also shown is the upper bound on miss detection probability given by~(\ref{eq:Chernoff}), that in this case (and for both the thresholds) can be approximated by the simple expression $\beta_n \le \exp(-\textnormal{SNR}\sqrt{2 \log n})$, after neglecting terms of higher order in $n$.

\subsection{Networks of random size}
\label{sec:randomsize}

The powerfulness of the theorems presented in the previous section allows us to consider the more general setting of network of random size $N$. Consider hence the following scenario. \\
$\bullet$ Sensors are randomly deployed in a two-dimensional region~${\cal A}$, according to a non-homogeneous Poisson field. The intensity function of this field is $\lambda(\bx)$, $\bx \in {\cal A}$, such that the average number of sensors in the region~${\cal A}$ is $\int_{\bx\in{\cal A}}\lambda(\bx)d\bx$. \\
$\bullet$ Some sensors are impaired before (or at) the act of communication. Thus, the number of active sensors $N$ is a subset of those globally available. The probability of a failure is unknown to the network, and is accordingly modeled as a random variable $Q$, independent of the deploying process. \\
$\bullet$ Conditioned on $Q=q$, the active sensors are selected independently and with probability $q$ from the total number of sensors available in~${\cal A}$. Given $Q=q$, the number of active sensors becomes a Poisson random variable with mean value $q\,\int_{\bx\in{\cal A}}\lambda(\bx)d\bx$. \\
$\bullet$ Accordingly, the average number of active sensors is
\[
\E(N)=
\E(Q)\,\int_{\bx\in{\cal A}}\lambda(\bx)d\bx\,
\dfz \nu.
\]
By introducing the normalized random variable $R={Q}/{\E(Q)}$, we have
\[
\Pr\left(
N=n|R=r 
\right)=
\frac{(\nu\, r)^n}{n!}\,e^{-\nu\, r}.
\]
$\bullet$
We focus on the asymptotic regime of increasingly large sensors, which in this context is formalized by $\nu \rightarrow \infty$.
From a practical perspective, note that an increasingly large value of $\nu$ may be due to an increasing large sensor density $\lambda(\bx)$, that corresponds to the asymptotic regime of a {\em dense} network (recall however that sensors' observations are iid), or to an increasingly large surveyed region~${\cal A}$, corresponding to the asymptotic regime of a {\em large} network. \\%While the physical meaning of varying $\nu$ changes accordingly depending on the particular applicative scenario, the overall effect on network performance is clearly unchanged. \\
$\bullet$
It is easy to show that\footnote{In fact, $\Pr\left(\left|\frac{N}{\nu}-r\right|>\epsilon|R=r\right)\rightarrow 0$ by the weak law of large numbers. Convergence in probability of the unconditioned random variable $N$ to $R$ easily follows by Lebesgue dominated convergence theorem.}
$N/\nu\rightarrow R$ in probability, when $\nu \rightarrow \infty$.

\begin{figure}
\centerline{\leavevmode \epsfxsize=3.5in %\epsfysize=1.2in 
\epsfbox{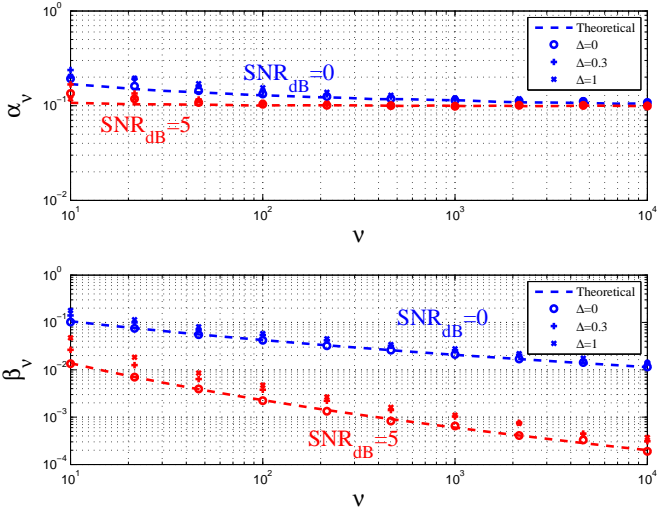}
}
\caption{$\ell$-MO transmission policy with networks of random size and Gaussian observation model. The values of $\alpha_n$ and $\beta_n$ are plotted, for different values of SNR$=(\theta_1+\theta_0)/\sigma$ and different values of $\Delta$. The curves labeled with ``theoretical'' are  shown for comparison and refers to $N$ deterministic, while those with $\Delta=0$ refer to the simple Poisson model with deterministic mean value.}
\label{fig:GaussComparison1}
\end{figure}

As an example, let us consider again the observation model in~(\ref{eq:gauss}) with $\ell$-MO transmission policy, but assume now that the effective network size $N$ is random according to the model described above, and suppose that $Q=\E(Q)+U$, with $U \sim \U(-\Delta/2,\Delta/2)$, where $\U(a,b)$ stems for the uniform distribution with support $(a,b)$. 
In Fig.~\ref{fig:GaussComparison1} the false alarm and miss detection probabilities, parametrized in SNR$=(\theta_1+\theta_0)/\sigma$, are shown for different values of $\Delta$, with $\E(Q)=0.5$. The curves are obtained by means of Monte Carlo computer experiments, except those labeled as ``theoretical''. These, plotted for comparison, refer to the case of $N=\nu$ deterministic and are obtained by numerical integration. 

We see that the false alarm probability $\alpha_n$ converges to its limiting value $\alpha=10^{-1}$, with a convergence rate that is faster for larger SNRs. The same is true for the convergence to zero of $\beta_n$ shown in the lower plot. It is also worth noting that the limit value of $\alpha_n$ is approached faster in the case of higher SNR, and almost at the same rate for $N$ random and $N$ deterministic. When the randomness grows, namely $\Delta$ becomes larger, we see that for the miss detection probability the convergence is slightly slowed down. 

\begin{figure}
\centerline{\leavevmode \epsfxsize=3.5in %\epsfysize=1.2in 
\epsfbox{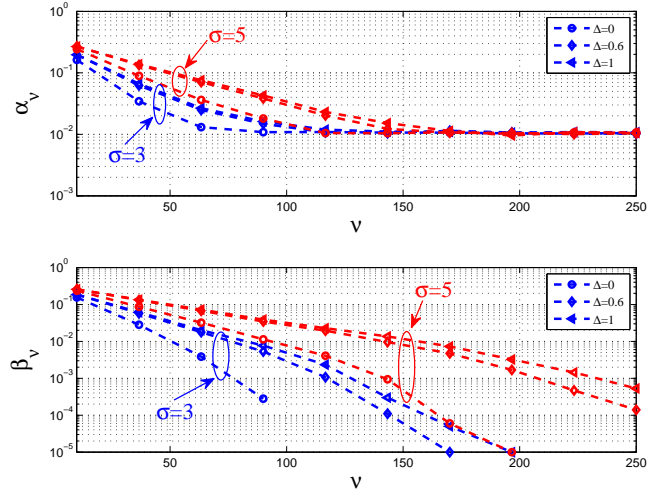}
}
\caption{Censoring transmission policy with networks of random size and non-Gaussian observation model. The values of $\alpha_n$ and $\beta_n$ are plotted, for two values of $\sigma$ and different values of the randomness index $\Delta$. The curves with $\Delta=0$ refer to the simple Poisson model with deterministic mean value.}
\label{fig:NonGauss2}
\end{figure}

\subsection{An example with censoring}
Consider again a network of random size, as described in Sect.~\ref{sec:randomsize}, but let us explore an example in which the transmission policy is based on a censoring strategy. Censoring techniques are commonly implemented in WSNs working under severe communication constraints, and amount to discard sensors' observations considered poorly informative for the detection purpose, see, e.g.,~\cite{rago-censoring,divpuc,tsitsikliscensoring,appadwedulaSP,addesso-censoring-SP07}. This can be obtained by selecting the transformation for the transmission policy according to the censoring rule:
\[
T(x)= \left \{
\begin{array}{ll}
x & \textnormal{ if } \; |x| \ge \theta_c \\
0 & \textnormal{ if } \; |x| < \theta_c 
\end{array}
\right .
\]
where $\theta_c>0$ is the censoring threshold.

To enrich the example, we adopt an observation model different from the Gaussian one considered so far. Specifically, assume that under ${\cal H}_1$ the sensors observe $X_i=W_i$ while under ${\cal H}_0$ they observe $X_i=-W_i$. The random variables $W_i$'s are iid with pdf
given by a mixture between a Gaussian and a Pareto density:
\[
p \frac{1}{\sqrt{2 \pi \sigma^2}} e^{-x^2/2 \sigma^2}
+ (1-p) \frac{b}{\theta}\left(\frac{x}{\theta}\right)^{-b-1} \, u(x-\theta)
\]
where $u(x)$ is the unit step function and $0<p<1$. In the computer experiments we set $\theta_c=\theta=b=1$, $p=0.5$, and $\E(Q)=0.7$. The detection errors $\alpha_n$ and $\beta_n$, computed by Monte Carlo simulations, are displayed in Fig.~\ref{fig:NonGauss2} for two values of $\sigma$ and different values of $\Delta$.
The general behavior is similar to that of Fig.~\ref{fig:GaussComparison1}: in particular the convergence is faster when there is less randomness in the system, as quantified by the value of~$\Delta$.

%%%%%%%%%%%%%%%%%%%%%%%%%%%%%%%%%%%%%%%%%%%%%%%%%%%
%%%%%%%%%%%%%%%%%%%%%%%%%%%%%%%%%%%%%%%%%%%%%%%%%%%
%\subsection{Randomly stopped acquisition stage}
\subsection{Observation-dependent network size}
%In this section we would like to consider a further scenario of interest for detection applications, where the presented strategies and results can be useful. A general, abstract model thereof can be formalized as follows.

The previous examples demonstrate the large versatility of the theorems provided in Sect.~\ref{sec:theory} that ensure the asymptotic  convergence of detection tests under a very broad class of applicative scenarios of practical relevance, including different transmission policies, different observation distributions, and very general network models. We now go even further by letting the random network size $N$ to be dependent upon the sensors' observations ---a possibility well encompassed in the theorems of Sect.~\ref{sec:theory}.

Let, as usual, the $i^{th}$ sensor of the network monitor the physical phenomenon of interest by collecting the sample $X_i$.
Suppose further that $X_i=S_i+W_i$, 
where the random variable $S_i$ models the intrinsic state of the observed phenomenon, while the random variable $W_i$ models the sensor measurement process; these two components are mutually independent and independent across sensors. 

As a distinct feature of this new scenario, we assume that the number of samples collected by the system is dependent upon the nature of the observed phenomenon, in such a way that the monitoring stage is ended at a certain random sample number\footnote{The random ``time'' defined in eq.~(\ref{eq:stoptime}) is by construction a Markov time, in that the event $N=n$ is determined by the observations of the first $n$ samples. Moreover, it will be a stopping time, provided that $\Pr(N<\infty)=1$.}:
\beq
N: \inf\left\{n: \varphi\left(S_1,S_2,\dots,S_n\right)>\nu\right\},\quad \varphi(\cdot)>0.
\label{eq:stoptime}
\eeq
To fix ideas, $\varphi(\cdot)$ might be thought as a measure of the energy emitted by the surveyed physical system, which is assumed to be limited. 
%and as such determines the time at which the physical phenomenon ends.
In our asymptotic framework, we are interested in increasingly large values of $N$, and this explains why the threshold in~(\ref{eq:stoptime}) has been just set to $\nu$. 

Consider for example a Gaussian shift-in-mean problem,
with $S_i \sim {\cal N}(-\theta_0,\sigma_s)$ under ${\cal H}_0$ and $S_i \sim {\cal N}(\theta_1,\sigma_s)$ under ${\cal H}_1$, 
%\[
%{\cal H}_0: \; S_i \sim {\cal N}(-\theta_0,\sigma) , \quad \textnormal{ } \quad 
%{\cal H}_1: \; S_i \sim {\cal N}(\theta_1,\sigma). 
%\]
%\beqa
%{\cal H}_0&:& \; S_i \sim {\cal N}(-\theta_0,\sigma_s) ,\nonumber\\
%{\cal H}_1&:& \; S_i \sim {\cal N}(\theta_1,\sigma_s),  \nonumber
%\eeqa
and with $W_i \sim {\cal N}(0,\sigma_w)$ under both hypotheses. Assume also that the stopping rule for the acquisition process is $N: \inf\left\{n: \sum_{i=1}^n S^2_i>\nu\right\},$
%\beq
%N: \inf\left\{n: \sum_{i=1}^n S^2_i>\nu\right\},
%\label{eq:stoptime2}
%\eeq
where $\varphi(s_1,s_2,\dots,s_n)=\sum_{i=1}^n s_i^2$ quantifies an energy expense (but for a normalization factor).
By the theory of renewal processes~\cite{ross-book}, we know, for $j=0,1$
%\beqa
%\frac{N}{\nu} &\rightarrow& \frac{1}{\theta_0^2+\sigma_s^2} \quad\textnormal{ under $\cH_0$}\nonumber
%\frac{N}{\nu}&\rightarrow& \frac{1}{\theta_1^2+\sigma_s^2} \quad\textnormal{ under $\cH_1$},
%\label{eq:Nconv}
%\eeqa
\beq
\frac{N}{\nu} \rightarrow \frac{1}{\theta_j^2+\sigma_s^2} \quad\textnormal{ under $\cH_j$}, 
\label{eq:Nconv0} 
\eeq
where the limit is to be intended with probability one.
Note further that the parameter $\nu$ is still related to the expected number of sensors. Indeed we also have~\cite{ross-book}, for $j=0,1$
\beq
\frac{\E(N)}{\nu} \rightarrow \frac{1}{\theta_j^2+\sigma_s^2} \quad\textnormal{ under $\cH_j$}. 
\label{eq:Nconv} 
\eeq

\begin{figure}
\centerline{\leavevmode \epsfxsize=3.5in %\epsfysize=1.2in 
\epsfbox{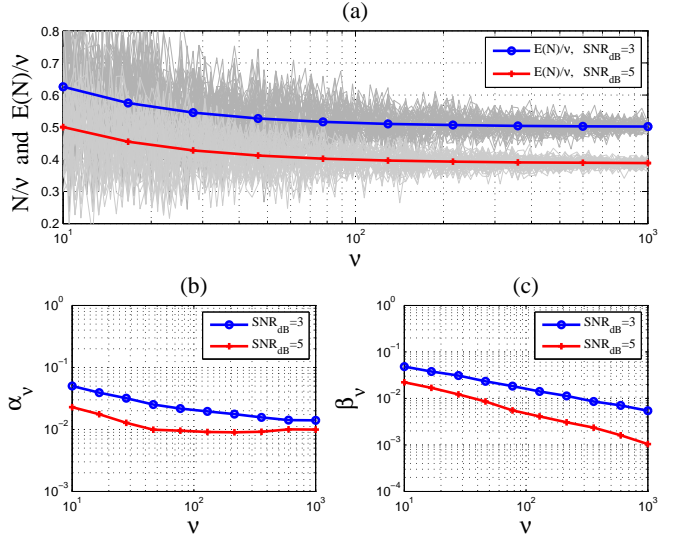}
}
\caption{$\ell$-MO transmission policy for a network whose size depends upon the observations. Top plot (valid for both $\cH_0$ and $\cH_1$) shows several realizations of $N/\nu$ (tiny curves) to illustrate the convergence with probability one in~(\ref{eq:Nconv0}), along with $\E(N)/\nu$ that converges to the same limit according to~(\ref{eq:Nconv}). Lower plots show the error probabilities for an asymptotic false alarm $\alpha=10^{-2}$, with threshold set by~(\ref{eq:gammaprime}). }
\label{fig:Stopping}
\end{figure}

We apply the proposed one-bit detection strategy to the above situation with $\theta_0=\theta_1$, $\sigma_s=\sigma_w=1$ and SNR$=(\theta_1+\theta_0)/\sqrt{\sigma_s^2+\sigma_w^2}$. The pertinent results are displayed in Fig.~\ref{fig:Stopping}, for an asymptotic false alarm probability $\alpha=10^{-2}$, with threshold set by eq.~(\ref{eq:gammaprime}) and using an $\ell$-MO policy.   
In Fig.~\ref{fig:Stopping} $(a)$ we show the convergence of the stopping number $N$ as given in~(\ref{eq:Nconv0}) and~(\ref{eq:Nconv}).
In Fig.~\ref{fig:Stopping} $(b)$ and $(c)$, we show the behavior of the error probabilities.

\subsection{Resilience to clock offset}
\label{sec:offset}

All the cases addressed above fall in the assumption of our theorems. As last example, we want to investigate briefly the robustness of the results with respect to models that slightly deviate from the formal assumptions of the theorems presented in Sect.~\ref{sec:theory}.
In particular, we have assumed so far that sensors are perfectly synchronized, and one should note that the transmission policy strictly relies on such assumption. What if time references of the sensors are slightly misaligned?
To make things simple, suppose that the clock of the generic sensor $i$ is perturbed by $U_i \sim \U(-\Delta_{clk}/2,\Delta_{clk}/2)$ that models the timing offset. In other words, sensor $i$ will attempt to transmit its local decision at the time instant
$1/\left|T(X_i)\right|+U_i$. Let us refer, for simplicity, to the Gaussian shift-in-mean example provided in Sect.~\ref{sec:gauss}. The dashed curves in Fig.~\ref{fig:nonparametric} show the effect of timing errors with different $\Delta_{clk}=0.1,1,2$, with respect to the nominal cases $\theta_0=\theta_1=1$ and $\theta_0=1.5,\theta_1=2$. 

As it can be seen, the test consistency seems to be preserved, thus evidencing a certain robustness of the proposed strategy.
On the other hand, and perhaps unsurprisingly, by increasing the offset error $\Delta_{clk}$, the performance worsen in the sense that the rate of convergence is slower, as consequence of the fact that the firing sensor may be different from the largest in modulus which conveys the largest information. Quantifying the effect of $\Delta_{clk}$ on the convergence rate and understanding whether a certain $\Delta_{clk}$ exists such that asymptotic convergence of the errors is lost, remain open problems.

%\begin{figure}
%\centerline{\leavevmode \epsfxsize=3in %\epsfysize=1.2in 
%\epsfbox{fig7.eps}
%}
%\caption{}
%\label{fig:offset1}
%\end{figure}

\section{Summary}
\label{sec:concl}

Distributed detection in large wireless sensor networks can be performed by the transmission of a single bit, exploiting the idea of ordered transmission policies. After casting such problem in a precise mathematical framework, we propose an easy-to-implement distributed statistical test whose asymptotic consistency is formally proved: Both the error probabilities can be controlled in the asymptotic regime of large network size, under a very broad class of observation models ---from classical Gaussian shift-in-mean to fairly more general measurement settings--- and applicative domains, including: nonparametric tests, likelihood-based transmission policies, censored systems, random network size, and observation-dependent sensor number.

\begin{appendices}

\section{Proof of Theorem 1}
\label{Sec:TH1}

We shall work under $\cH_1$, thus proving eqs.~(\ref{eq:mainth_H1}), and consistently skip the explicit dependence upon the hypothesis for notational ease. The proof of eqs.~(\ref{eq:mainth_H0}) follows straightforwardly.
Let us introduce the sequence of events $\cE_{\nu}=\left\{\cPMM \geq \cNMM   \right\}$. In view of the definition of the detection statistic $\cM_N $, the claim of the theorem will be certainly true if $\Pr\left( \cE_{\nu} \right) \rightarrow 1$. In order to show that this convergence actually takes place, let us elaborate as follows.
Let $0<x_0<x_1<+\infty$ such that
\beq
\Pr(R\in [x_0,x_1])\geq 1-\epsilon,
\label{eq:partition}
\eeq
where, we recall, $N/\nu$ converges in probability to $R$.
We are now legitimate to write
\beqa
\Pr\left( \cE_{\nu} \right) &\geq&
\Pr\left( \cPMM \geq \cNMM  , \frac{N}{\nu}\in [x_0,x_1] \right)\nonumber\\
&\geq&
\Pr\left( \cM^{+}_{n_0} \geq \cM^{-}_{n_1}  , \frac{N}{\nu}\in [x_0,x_1] \right)
%\nonumber\\
\label{eq:primogenita}
\eeqa
where we have defined $n_j=\lfloor{\nu\,x_j}\rfloor$, $j=0,1$, and the last inequality follows by obvious properties of maxima and minima.
Assume for now that 
\beq
\lim_{\nu\rightarrow\infty} \Pr\left( \cM^{+}_{n_0} \geq \cM^{-}_{n_1}\right)=1.
\label{eq:secondogenita}
\eeq
This would imply that the last limit in eq.~(\ref{eq:primogenita}) equals
\[
\lim_{\nu\rightarrow\infty}
\Pr\left(\frac{N}{\nu}\in [x_0,x_1] \right)
=\Pr\left(R\in [x_0,x_1] \right)
\geq 1-\epsilon
\]
where the inequality follows by eq.~(\ref{eq:partition}). Inequality~(\ref{eq:primogenita}), $\epsilon$ being arbitrary, implies 
$\liminf_{\nu\rightarrow\infty}\Pr(\cE_\nu)=1$,
and hence $\Pr(\cE_\nu)\rightarrow 1$ as $\nu$ diverges. 

It remains thus to show that eq.~(\ref{eq:secondogenita}) holds.
To this aim, it is expedient to work in terms of the normalized variables 
%\[
%\widetilde{\cM}^{+}_{n_0}=\frac{\cM^{+}_{n_0}-b^{+}_{n_1}}{a^{+}_{n_1}},\quad 
%\widetilde{\cM}^{-}_{n_1}=\frac{\cM^{-}_{n_1}-b^{+}_{n_1}}{a^{+}_{n_1}}.
%\]
$\widetilde{\cM}^{+}_{n_0}=(\cM^{+}_{n_0}-b^{+}_{n_1})/a^{+}_{n_1}$ and 
$\widetilde{\cM}^{-}_{n_1}=(\cM^{-}_{n_1}-b^{+}_{n_1})/a^{+}_{n_1}$.
Note first that, by assumption $ii)$, it is straightforward to conclude that 
\beq
\lim_{\nu\rightarrow\infty}
\Pr\left(\widetilde{\cM}^{+}_{n_0}\leq x\right)
=\left(G^+(x)\right)^{\eta}, \quad \textnormal{where} \quad \eta=\frac{x_0}{x_1} .
\label{eq:newConv}
\eeq
Furthermore, by assumption $iv)$ we know that the $Z_i$'s are right-tail dominant under $\cH_1$, which implies (see~\cite{Resnick71}, proof of Theorem~2.1), for all $x$ with $G^{+}(x)\neq 0$ and $G^{+}(x)\neq 1$, that
\beq
\Pr\left(\widetilde{\cM}^{-}_{n_1} < x \right)  \rightarrow 1.
\label{eq:TailDomImplic}
\eeq    

It is convenient to study separately the different admissible attraction domains.   
Let us first consider the case that $G^+(x)$ is a Gumbel distribution. We have
\beqa
\lefteqn{\Pr\left( \widetilde{\cM}^{+}_{n_0} \geq \widetilde{\cM}^{-}_{n_1} \right)}   \nonumber\\
&\geq&\Pr\left( \widetilde{\cM}^{+}_{n_0} - \widetilde{\cM}^{-}_{n_1} \geq 0 , \widetilde{\cM}^{-}_{n_1}<-\delta \right)\nonumber\\
&\geq&\Pr\left( \widetilde{\cM}^{+}_{n_0} + \delta \geq 0 , \widetilde{\cM}^{-}_{n_1}<-\delta\right)
\label{eq:boundbound}
\eeqa
where $\delta>0$ is arbitrarily large. 
On the other hand, eq.~(\ref{eq:TailDomImplic}), along with eq.~(\ref{eq:newConv}) implies
\beqa
\lim_{\nu\rightarrow \infty} 
\Pr\left( \widetilde{\cM}^{+}_{n_0} + \delta \geq 0 , \widetilde{\cM}^{-}_{n_1}<-\delta\right)=
1-e^{-\eta\,e^{\delta}},
\eeqa
yielding, in the light of eq.~(\ref{eq:boundbound}) and being $\delta$ arbitrary,
\[
\liminf_{\nu\rightarrow\infty}
\Pr\left( \widetilde{\cM}^{+}_{n_0} \geq \widetilde{\cM}^{-}_{n_1} \right)
=1.
\]

Let us switch now to the case that $G^+(x)$ is Fr\'echet distributed.
We first note that $\wcNM$ now vanishes in probability.
Indeed, thanks to eq.~(\ref{eq:TailDomImplic}), we have
$\Pr\left(\widetilde{\cM}^{-}_{n_1} >\epsilon \right)\rightarrow 0$
and
\[
\Pr\left(\widetilde{\cM}^{-}_{n_1} < -\epsilon \right)\leq
\Pr\left(\widetilde{\cM}^{-}_{n_1} < 0 \right)=(1-F_Z(0))^{n_1}\rightarrow 0,
\]
having used the fact that, for the Fr\'echet domain of attraction, $b^+_n=0$. 
Moreover, by eq.~(\ref{eq:newConv}), the sequence $\widetilde{\cM}^{+}_{n_0}$ converges to a Fr\'echet random variable.
Slutsky's theorem allows to conclude that the sequence
$\widetilde{\cM}^{+}_{n_0}-\widetilde{\cM}^{-}_{n_1}$
converges in distribution to a non-negative random variable, implying the desired result~(\ref{eq:secondogenita}).~\hfill$\bullet$

\section{Proof of Theorem 2}
\label{Sec:TH2}
Let's start with part $i)$, and accordingly consider the term 
\beq
\beta_\nu=\Pr\left(\cM_N<0;\cH_1\right).
\label{eq:betagamma0}
\eeq
Now, for the case that $G^+(x)$ is Gumbel, by the convergence in distribution of $(\cM_N-b^+_\nu)/a^+_\nu$, and the divergence of the term $b^+_\nu/a^+_\nu$, we desume that $\cM_N/b^+_\nu\rightarrow 1$ in probability, implying $\beta_\nu\rightarrow 0$ in the light of eq.~(\ref{eq:betagamma0}).
For the case that $G^+(x)$ if Fr\'echet, we know that $\cM_N/a^+_\nu$ converges in distribution to $H^+(x)$, which is supported 
on $x>0$, and again $\beta_\nu\rightarrow 0$. Similar reasoning will lead to $\alpha_\nu\rightarrow 0$.

Let us switch to the part $ii)$, and consider first $\gamma_{\nu}$ as in eq.~(\ref{eq:gamma}). 
For this case convergence of $\alpha_{\nu}$ toward $\alpha$ is nothing but eq.~(\ref{eq:mainth2}).
Let us move to $\gamma_\nu$ defined as in eq.~(\ref{eq:gammaprime}).
The attraction properties of $\cNM$ imply
\[
\Pr\left(\frac{\cNM-b^-_n}{a^-_n}<-\gamma;\cH_0\right)\rightarrow G^-(-\gamma)=\widetilde\alpha
\]
where the last equality follows by eq.~(\ref{eq:alphadet}).
On the other hand, by the definition of the refined threshold $\gamma_n$ in eq.~(\ref{eq:gammaprime})
\[
\Pr\left(\frac{\cNM-b^-_n}{a^-_n}<-\frac{\gamma_n+b^-_n}{a^-_n};\cH_0\right)=\widetilde\alpha.
\]
In view of Lemma 11.2.1 in~\cite{Lehmann-testing}, this allows concluding that
$(\gamma_n+b^-_n)/a^-_n\rightarrow\gamma$. By eq.~(\ref{eq:mainth2}) we have thus
\beqa
\alpha_\nu&=&\Pr\left(\cM_N>\gamma_\nu;\cH_0\right)\nonumber\\
&=&\Pr\left(\frac{\cM_N+b^-_{\nu}}{a^-_{\nu}}>\frac{\gamma_\nu+b^-_\nu}{a^-_\nu};\cH_0\right)
\rightarrow \E\left(\widetilde\alpha^{R_0}\right)=\alpha,\nonumber
\eeqa
the last equality following by eq.~(\ref{eq:inversion}).

Let us now switch to the analysis of $\beta_{\nu}$. 
Note that the threshold $\gamma_{\nu}$ in eq.~(\ref{eq:gamma}) is negative, at least for sufficiently large $\nu$.
Indeed, if $G^-(x)$ is Gumbel, $b^-_{\nu}/a^-_{\nu}\rightarrow\infty$ and $a^-_{\nu}>0$; 
if $G^-(x)$ is Fr\'echet, $\gamma<0$, being the support of $1-H^-(x)$ confined to the negative axis, see eq.~(\ref{eq:Hpmdef}).
Simple inspection show that the threshold $\gamma_{\nu}$ in eq.~(\ref{eq:gammaprime}) is as well negative, at least for sufficiently large $\nu$.
Thus, for sufficiently large $\nu$ and for both choices of the thresholds one can write  
$\beta_\nu\leq\Pr\left(\cM_N<0;\cH_1\right)$, and the proof is now complete.~\hfill$\bullet$

\section{Proof of Theorem 3}
\label{Sec:TH3}

Let us first check the validity of condition $iv)$ in Theorem 1. 
The symmetry of the function $\phi(x)$ imply, for the considered shift-in-mean problem, 
$f_{Z}(x ;\cH_1)=f_{Z}(-x ;\cH_0)$. 
The well-known nesting rule for the log-likelihoods~\cite{vantrees-book1} further gives
$\log\frac{f_{Z}(x ;\cH_1)}{f_{Z}(x ;\cH_0)}=x$.
Combining the above results gives
$
f_{Z}(x ;\cH_0)=f_{Z}(-x ;\cH_0)e^{-x}$, and $f_{Z}(x ;\cH_1)=f_{Z}(-x ;\cH_1)e^{x}$,
which clearly implies that $Z$ is right-tail dominant under $\cH_1$ and left-tail dominant under $\cH_0$. 

Let us now prove eq.~(\ref{eq:Chernoff}). 
To this aim, we write the log-likelihood ratio of $\cM_{N}$
\beqa
&&\log \frac
{ \sum_{n=1}^{\infty} f_{\cM_n} \left(x ;\cH_1 \right) \Pr \left(N=n\right)}
{ \sum_{n=1}^{\infty} f_{\cM_n} \left(x ;\cH_0 \right) \Pr \left(N=n\right)}
\nonumber\\
&&= \log \frac
{f_{Z}\left(x ;\cH_1\right)\, \sum_{n=1}^{\infty} n\,h^{n-1}_1(x)  \Pr \left(N=n\right)}
{f_{Z}\left(x ;\cH_0\right)\, \sum_{n=1}^{\infty} n\, h^{n-1}_0(x) \Pr \left(N=n\right) }
\nonumber\\
&&=x+
\log \frac
{ \sum_{n=1}^{\infty} n\,h^{n-1}_1(x)  \Pr \left(N=n\right)}
{ \sum_{n=1}^{\infty} n\, h^{n-1}_0(x) \Pr \left(N=n\right) }
\nonumber
\eeqa
where in the last equality we again applied the nesting rule.
Moreover, it is easy to check that, in the shift-in-mean case with even $\phi(x)$, we have $h_1(x)=h_0(x)$, finally yielding
$
\log \frac{f_{\cM_{N}} \left(x ; \cH_1 \right) }{f_{\cM_{N}} \left(x; \cH_0 \right)}=x.
$
At this point we are legitimate to use the Chernoff bound:   
$\beta_{\nu}= 
\Pr \left(\cM_{N} < \gamma_{\nu}; \cH_1 \right) 
\leq e^{\gamma_{\nu}}$,
and the proof is complete.~\hfill$\bullet$ 

\end{appendices}

\vspace*{-5pt}

%\bibliographystyle{IEEEtran}
%\bibliography{IEEEabrv,../../bib/mybib}
%%\bibliography{mybib}
%\end{document}


\end{document}